# Dark trions govern the temperature-dependent optical absorption and emission of doped atomically thin semiconductors

Ashish Arora[†,*], Nils Kolja Wessling[†], Thorsten Deilmann[¶], Till Reichenauer[†], Paul Steeger[†], Piotr Kossacki[§], Marek Potemski[§,‡], Steffen Michaelis de Vasconcellos[†], Michael Rohlfing[¶], and Rudolf Bratschitsch[†,*]

[†]Institute of Physics and Center for Nanotechnology, University of Münster, Wilhelm-Klemm-Straße 10, 48149 Münster, Germany
[¶]Institute of Solid State Theory, Wilhelm-Klemm-Straße 10, University of Münster, 48149 Münster, Germany
[§]Institute of Experimental Physics, Faculty of Physics, University of Warsaw, ul. Pasteura 5, 02-093 Warszawa, Poland
[‡]Laboratoire National des Champs Magnétiques Intenses, CNRS-UGA-UPS-INSA-EMFL, 25 avenue des Martyrs, 38042 Grenoble, France
*Email: arora@uni-muenster.de, Rudolf.Bratschitsch@uni-muenster.de

**We perform absorption and photoluminescence spectroscopy of trions in hBN-encapsulated WSe$_2$, WS$_2$, MoSe$_2$, and MoS$_2$ monolayers, depending on temperature. The different trends for W- and Mo-based materials are excellently reproduced considering a Fermi-Dirac distribution of bright and dark trions. We find a dark trion, $X_D^-$ 19 meV *below* the lowest bright trion, $X_1^-$ in WSe$_2$ and WS$_2$. In MoSe$_2$, $X_D^-$ lies 6 meV *above* $X_1^-$, while $X_D^-$ and $X_1^-$ almost coincide in MoS$_2$. Our results agree with *GW*-BSE *ab-initio* calculations and quantitatively explain the optical response of doped monolayers with temperature.**

**Introduction.** Everyday optoelectronic devices such as solar cells, photodiodes, light emitting diodes, and lasers critically depend on the knowledge about the energy level structure of the semiconducting materials. In particular, the emission of light strongly depends on optical selection rules governing the lowest-energy optical transitions. Recently, it has been found that the atomically thin semiconducting transition metal dichalcogenides (TMDCs) of the form W$X_2$ ($X$ = S, Se) or Mo$X_2$ ($X$ = S, Se, Te) are fundamentally different from the optical point of view. Despite their similar crystal structure, the photoluminescence (PL) of neutral excitons in monolayer MoSe$_2$ [1–3], MoS$_2$ [3,4] and MoTe$_2$ [5] decreases from cryogenic up to room temperature, while the PL increases for monolayer WSe$_2$ [2–4,6] and WS$_2$ [3,7]. This behavior has been qualitatively explained by the presence of dark excitons. Due to the uniquely contrasting spin character of the conduction bands [1,4,6,8,9], the lowest neutral exciton transition in W$X_2$ is optically disallowed ("dark"), and it is dipole-allowed ("bright") in Mo$X_2$ [1,5,15–18,6,8–14]. Besides these spin-forbidden dark excitons, momentum-forbidden dark excitons with electron and hole residing in different valleys play a role [19,20].

The optical response of doped TMDC monolayers is governed by trions [9,21–23]. Similar to neutral excitons, bright [11,21,24] and dark [11] trions exist. However, their role in the temperature-dependent absorption and photoluminescence emission is poorly understood. In particular, dark trions are notoriously difficult to study in experiments [13,14,25,26]. Since doped atomically thin semiconductors hold the promise for novel 'valleytronic' devices [20,27], it is of utmost importance to gain a quantitative understanding of bright and dark trions.

We measure the optical absorption and PL emission of trions in TMDC monolayers of WSe$_2$, WS$_2$, MoSe$_2$, and MoS$_2$ encapsulated in hBN as a function of temperature. We reveal that the combined effects of absorption transfer between trions and excitons, and the thermal distribution of trions between the bright and dark states leads to the unique PL response of the different 2D semiconductors.

**Bright and dark trions.** Due to strong spin-orbit coupling, the conduction and valence bands at the K point of the Brillouin zone of TMDCs are almost completely polarized. The transitions between bands with opposite spin have vanishing oscillator strength (dark excitons/trions). Assuming a negative doping, four different types of excitations are possible: neutral bright excitons (X$^0$), neutral dark excitons (X$_D$), bright trions (X$^-$), and dark trions (X$_D^-$) [11]. Fig. S1 (see Supplemental material [28], which includes Refs. [29–33]) depicts a k-space representation of these excitations and quantitative results of our *ab-initio* calculations employing the *GW*+BSE approach with its extension to trions [11,34–37]. In our calculation of W$X_2$ monolayers in vacuum, $X_1^-$ is the lowest-bright state, while the dark trion $X_D^-$ lies about 30 meV below $X_1^-$. In contrast, for MoS$_2$ and MoSe$_2$, the dark trion $X_D^-$ resides only 5 meV and 10 meV below the bright trion $X_1^-$, respectively. Because the relative accuracy of our method is about 10 meV, bright and the dark trions in Mo-based materials are almost energetically degenerate on this energy scale. The effect of surrounding hBN on the relative separation between bright and dark trions (when compared to vacuum around the monolayer) is less than 10 meV [38], which is within our numerical accuracy.

**Temperature-dependent absorption and emission spectra.** Figure 1(a) and 1(e) present optical absorption and PL spectra of a hBN-encapsulated WSe$_2$ monolayer, measured for temperatures $T$ = 5 - 300 K (also Fig. S3 [28]). At cryogenic temperatures, narrow resonances with nearly Lorentzian line shapes corresponding to the two trions ($X_1^-$ and $X_2^-$) and the neutral exciton X$^0$ are clearly resolved. The presence of two trions points towards a negative doping of the WSe$_2$ monolayer [11,24,39]. Narrow linewidths (Table S1 [28]) approaching the homogeneous regime confirm the high quality of our sample [40–42]. Figure 1(b) – (d) depict optical absorption and Fig. 1(f) – (h) show PL spectra at different temperatures for hBN-encapsulated 1L WS$_2$, MoSe$_2$, and MoS$_2$. Similar to WSe$_2$, WS$_2$ also exhibits two trions and one neutral exciton. For 1L MoS$_2$, two trions are clearly discernible (Fig. 1(d)), in agreement with recent reports [43]. For 1L MoSe$_2$, only one trion is visible (Fig. 1(c)). Theoretically, three nearly degenerate bright trion resonances are expected for MoSe$_2$, which possibly are not resolved in the experiment [11]. The PL lines $U^1$ and $U^2$ in Fig. 1(e) and 1(f) have previously been associated either with the emission from defect-bound excitons [44] or with dark excitons coupled to chiral phonons [45]. In agreement with previous works, their intensity strongly reduces with rising temperature from 5 K to 75 K [6,44].

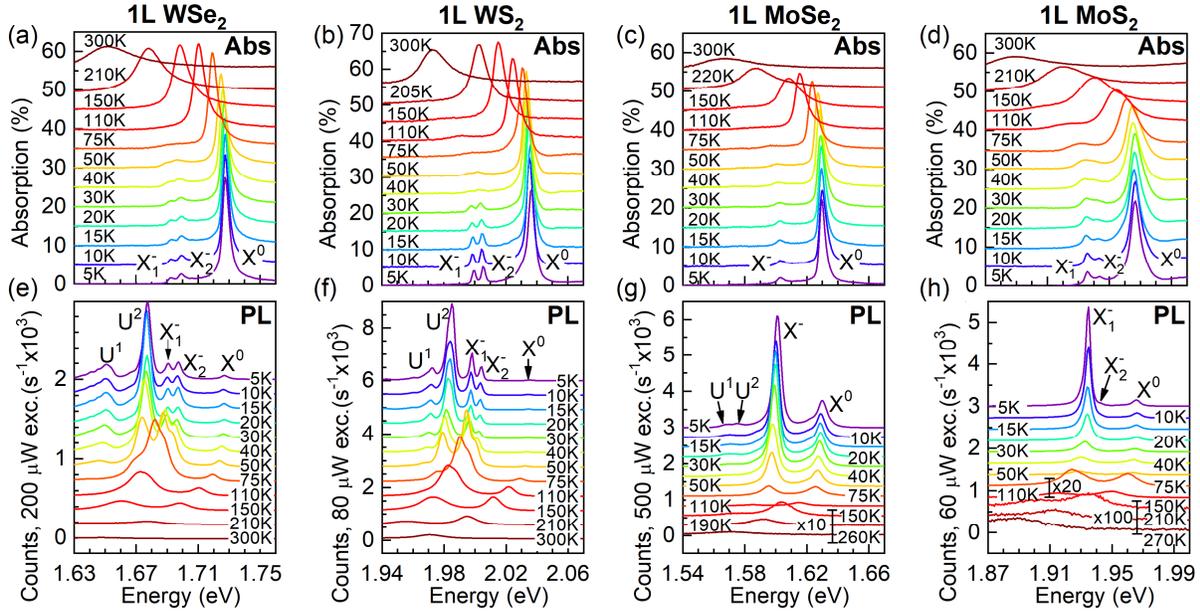

**Figure 1.** Optical absorption spectra of hBN-encapsulated (a) WSe$_2$, (b) WS$_2$, (c) MoSe$_2$, and (d) MoS$_2$ monolayers on sapphire substrate, measured as a function of temperature $T = 5 - 300$ K. (e) to (h) photoluminescence (PL) spectra as a function of temperature. The spectra are vertically shifted for clarity. Low PL intensities in (g) and (h) are amplified by factors mentioned with the spectra.

**Absorption transfer between trions and excitons.** To gain a comprehensive understanding of the optical response of the 2D semiconductors at different temperatures, we fit the trion and neutral exciton resonances in the absorption and PL spectra of Fig. 1(a)-(h) with normalized Lorentzians, as described in Ref. [28] and Fig. S9 therein. The area under the absorption lines (integrated absorption) of trions $\alpha_{X^-}(T)$ is plotted in Figs. 2(a – d) as a function of temperature. $\alpha_{X^-}(T)$ is given as a product of the number density of the optically excited trions and their oscillator strength [46,47]. In all cases, the integrated absorption of the trions decreases as the temperature rises, while it increases for the neutral excitons (see Fig. S4 [28]). This observation is in agreement with previous temperature-dependent spectroscopy of trion-exciton pairs in II-VI quantum wells [46–48], 1L WSe$_2$ [6], and 1L WS$_2$ [49], where a relationship between the excess carrier density and the integrated absorption of the trions was found. Such a connection has also been established in magneto-reflectance and magneto-PL spectroscopy of III-V and II-VI quantum wells [47,50,51] and TMDC monolayers [52]. The mechanism of absorption transfer between trions and excitons with temperature is explained as follows.

In a doped semiconductor (considering negative doping henceforth) at thermal equilibrium, excess charges occupy phase space around the band extrema of the Brillouin zone following a Fermi-Dirac (F-D) distribution, $F_e(T)$ (Fig. 3(a)).

The distribution function of trions (and therefore the probability of trion creation) is related to the one of the excess electrons through the mass ratio i.e. $F_{X^-}(T) = F_e\left(T \frac{M^*_{X^-}}{m^*_e}\right)$ [46,53], where $m^*_e$ and $M^*_{X^-} = 2m^*_e + m^*_h$ are the electron and trion effective masses, $m^*_h$ is the hole effective mass (see Figs. S5(a) and (b) [28]). The formation of trions by optical excitation (within the light cone), is favored at low temperatures, resulting in a large integrated absorption of the trion. The creation of neutral excitons is quenched due to Pauli blocking, as well as by screening in the presence of free carriers [54]. As temperature increases, the electrons redistribute towards higher energies and away from the light cone (Fig. 3(a)). This process reduces the number of optically active trions [36], leading to a reduction of their integrated absorption (see Fig. S5(c)-(f) [28]). Simultaneously, the formation of neutral excitons is enhanced due to an increased number of unoccupied states around the K points, leading to an increase of exciton integrated absorption (see Fig. S4 for the experimental data for the four materials). The measured integrated absorption of the trions in Figs. 2(a) to 2(d) for the four materials is fitted in the entire temperature range using Eq. S4 [28] following the physical picture in Fig. S5. In the next section, these functions are used for calculating the temperature-dependent integrated PL intensity (area under a PL peak, or 'PL intensity' henceforth) of the trions.

**Temperature dependence of the trion photoluminescence.** For WSe$_2$, the PL intensity of both trions steadily increases up to $T = 70$ K, followed by a decrease (Fig. 2(e)). For WS$_2$, the PL intensity of both trions shows an initial reduction up to $T = 30$ K, also followed by an increase up to $T = 70$ K, and a decrease at higher temperatures (Fig. 2(f)). Figures 2(g) and 2(h) depict the measured trion emission intensities of MoSe$_2$ and MoS$_2$. They strongly differ from the W-based materials,

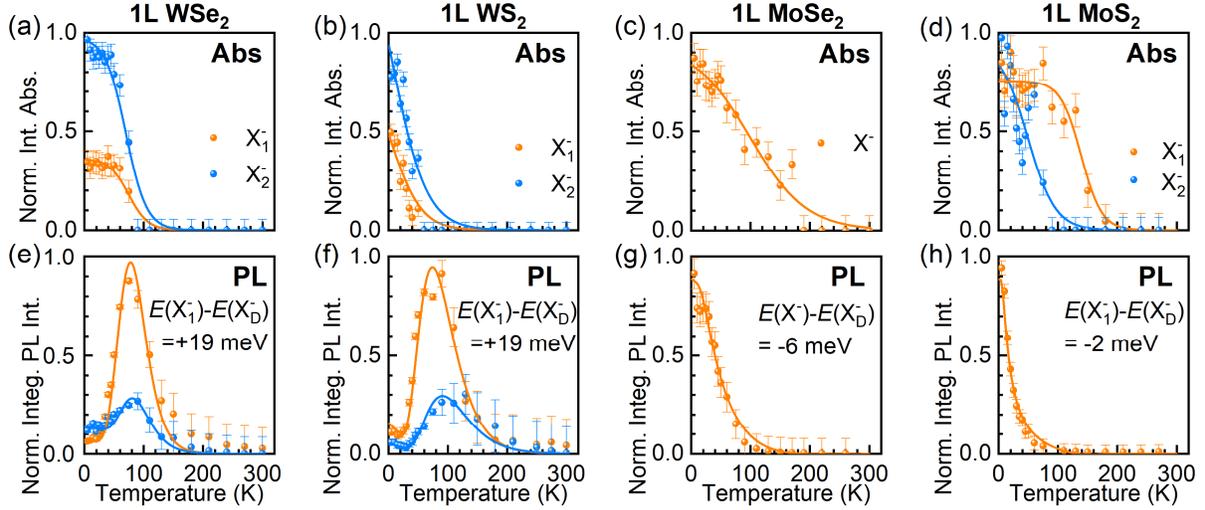

**Figure 2.** Normalized integrated absorption of the trions (orange and blue spheres) deduced from the measured absorption spectra as a function of temperature for hBN-encapsulated (a) WSe$_2$, (b) WS$_2$, (c) MoSe$_2$, and (d) MoS$_2$ monolayers. Solid lines are the fitted curves using Eq. S4 [28], as explained in the main text. (e) to (h) show the normalized integrated photoluminescence (PL) intensities of the trions obtained from fitting the measured PL spectra as a function of temperature for the four materials. Solid lines are the fitted curves using Eq. 1 as explained in the main text.

because, here, the emission intensity decreases with rising temperature.

The striking behavior of the two W-based materials is explained as follows. In the simplest approximation, the trion PL intensity is proportional to the density of the optically active trions $n_{X^-}$ (at thermal equilibrium) around the K points of the Brillouin zone within the light cone and their integrated absorption ($\alpha_{X^-}(T)$):

$$I_{X^-}(T) = n_{X^-}(T)\, \alpha_{X^-}(T)$$

$$= \left( n_{bg} + \frac{n_0}{1+\exp\left(\frac{E(X^-)-E(X_D^-)}{k_B T}\right)} \right) \alpha_{X^-}(T). \quad (1)$$

Here, $n_0$ is the total (bright and dark) density of trions, $E(X^-)$ and $E(X_D^-)$ are the energies of the bright and the dark trion, $n_{bg}\alpha_{X^-}(T)$ is the PL background, representing a possible PL increase due to optical doping or a PL decrease due to non-radiative processes, and $k_B$ is the Boltzmann constant. Unlike neutral excitons, an extra electron is required for the creation of negative trions. Therefore, the population of trions in quasithermal equilibrium at low excitation densities is directly proportional to the density of excess electrons. To recombine and emit a photon, a trion should be located within the light cone to satisfy the linear momentum conservation. In addition, the three carriers constituting a trion need to have the correct spin orientation to be optically bright (Fig. S1 [28]). In steady state, one can assume a F-D distribution of the bright and dark trions. In this case, the trion luminescence from the bright states is governed by F-D statistics. In first approximation, neutral excitons, being bosons, do not directly participate in this process. In W-based materials, the lowest energy trion state is optically dark (Fig. S1(c) [28]) [1,4,6,8,9,11,12,14,15,55], leading to an inefficient PL emission (Fig. 3(b)) at low temperatures. When the temperature is initially increased (5 K < $T$ < 30 K), the integrated absorption of WS$_2$ drops sharply (Fig. 2(b)), while it decreases only slightly for WSe$_2$ (Fig. 2(a)). As a consequence, the trion PL intensity of WS$_2$ falls initially, while it shows a slight increase for WSe$_2$. With rising temperature (30 K < $T$ < 70 K), the bright trion state with energy $E(X^-)$ is increasingly populated following a F-D distribution, resulting in a stronger PL (Fig. 3(b), right panel) for both materials. However, as the integrated absorption of trions drastically decreases at higher temperatures $T > 70$ K

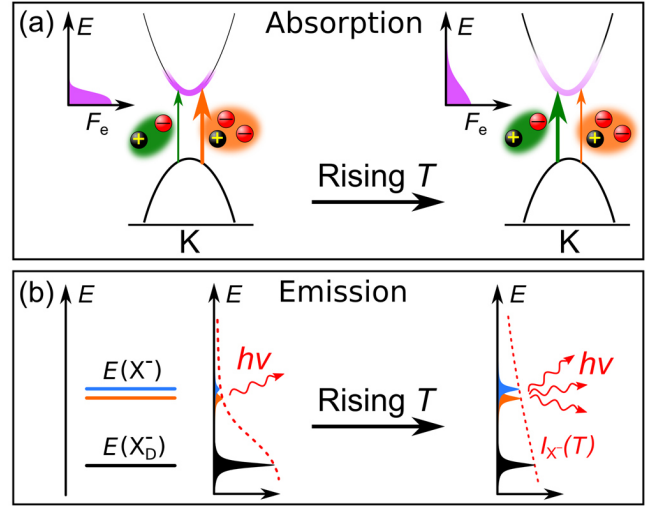

**Figure 3.** (a) Schematic drawing visualizing the transfer of absorption between neutral excitons (green vertical arrow) and trions (orange vertical arrow) with rising temperature due to the thermal redistribution of free carriers (purple) in momentum space. A thicker vertical arrow represents a larger transition strength. It reduces for trions with rising temperature. (b) Schematic drawing visualizing the increasing emission intensity of the bright trion when the temperature rises for W-based TMDC monolayers, as explained in the main text.

**Table 1. Energies of the dark trion $X_D^-$, bright trions ($X_1^-$ and $X_2^-$) and the bright exciton $X^0$ in the four materials. The last two columns show the energy difference of $X_D^-$, with respect to the $X_1^-$, in experiment and *GW*-BSE *ab-initio* calculations.**

| Monolayer (hBN encapsulated) | Transition energy (meV) | | | | $E(X_1^-) - E(X_D^-)$ (meV) | |
|---|---|---|---|---|---|---|
| | $X_D^-$ (Exp.) | $X_1^-$ (Exp.) | $X_2^-$ (Exp.) | $X^0$ (Exp.) | Exp. | Th. |
| WSe$_2$ | 1677±3 | 1691±1 | 1697±1 | 1727±1 | +19±3 | +30±10 |
| WS$_2$ | 1981±3 | 1998±1 | 2004±1 | 2035±1 | +19±3 | +30±10 |
| MoSe$_2$ | 1609±3 | 1603±1 | - | 1630±1 | -6±3 | +5±10 |
| MoS$_2$ | 1938±3 | 1936±1 | 1943±1 | 1966±1 | -2±3 | +10±10 |

(Figs. 2(a) and 2(b)), the PL intensity decreases again for both W-based materials.

We use Eq. (1) to fit the trion PL intensities of WSe$_2$ and WS$_2$ in Figs. 2(e) and 2(f) (solid lines). The energies of the two bright trions $E(X_1^-)$ and $E(X_2^-)$ are fixed to the values obtained from the measured PL spectra, and an additional dark-trion state of energy $E(X_D^-)$ is assumed. The two trion-intensity curves are collectively fit by only using PL data for $T \leq 150$ K, so that the role of non-radiative processes and phonon-absorption-related processes on the relative positions of the bright and dark trions at higher temperatures is minimized. From our fitting (Table 1), we find that the additional (dark) trion state $X_D^-$ is located $19 \pm 3$ meV below the lowest bright trion $X_1^-$ both in WSe$_2$ and WS$_2$ (Fig. 4). This value is in excellent agreement with gate-voltage-dependent PL studies in hBN-encapsulated WSe$_2$ (20 meV [25] and 22 meV [56]), and a very recent magneto-photoluminescence study on hBN-encapsulated WS$_2$ (19 meV) [26]. We emphasize that we obtain the same value of $E(X_D^-)$ within our error bars for both materials, when applying our model to each bright trion independently (unlike fitting the two trions collectively above, where a common $E(X_D^-)$ was assumed).

For MoS$_2$, only $X_1^-$ is used in our model, since $X_2^-$ is very weak in PL and cannot be tracked at higher temperatures. We find a dark trion state located $6 \pm 3$ meV and $2 \pm 3$ meV *above* the bright trion state for MoSe$_2$ and MoS$_2$, respectively, elucidating the "optically-bright nature" of Mo-based materials (Fig. 4). Importantly, it is not possible to model the experimental data, if the dark trion states are neglected in the two bright materials i.e. MoSe$_2$ and MoS$_2$.

**Nature of dark trion states.** For dark transitions, two mechanisms have to be considered: (a) spin-forbidden transitions, and (b) momentum-indirect transitions [20]. Our experiments point towards spin-forbidden dark excitons. Momentum-dark states can show photoluminescence at rising temperatures because of phonon-assisted recombination processes. Thus a PL line emerges or is enhanced [57,58]. However, we do not observe this behavior in our experiment. Furthermore, transitions from dark to bright trions would need to absorb phonons in case of momentum-dark transitions and thus would distinctly deviate from a Fermi-Dirac distribution, which is contrary to the experimental results.

In conclusion, we observe a strongly different behavior of the photoluminescence emission of trions in WSe$_2$, WS$_2$ vs. MoSe$_2$, and MoS$_2$ monolayers as a function of temperature. Our experimental results are excellently modelled by a dark trion state energetically below the bright trions in WSe$_2$ and WS$_2$, and by a dark trion state almost equal in energy with the bright trion in MoSe$_2$ and MoS$_2$. These results are also in good agreement with our *ab-initio* calculations based on *GW*-BSE theory. Our work provides a comprehensive understanding of the radiative phenomena in doped TMDCs as a function of temperature, and is important for a new generation of valleytronic devices involving the creation and readout of long-lived dark trions.

AUTHOR INFORMATION

**Funding Sources:** The authors acknowledge financial support from the German Research Foundation (DFG projects no. AR 1128/1-1, AR 1128/1-2, and DE 2749/2-1). MP and PK were supported by the ATOMOPTO project carried out within the TEAM programme of the Foundation for Polish Science co-financed by the European Union under the European Regional Development Fund. MP also acknowledges the EU Graphene Flagship project (No. 785219). The computing time granted by the Paderborn Center for Parallel Computing (PC2) is gratefully acknowledged.

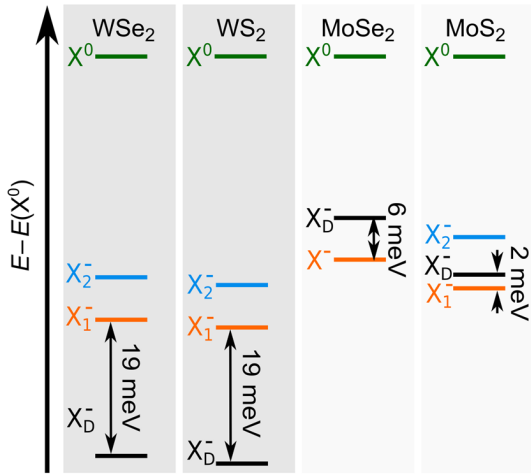

**Figure 4.** Relative energies of the bright trions ($X_1^-$ and $X_2^-$) and the dark trion $X_D^-$ with respect to the neutral exciton $X^0$ for hBN-encapsulated monolayers of WSe$_2$, WS$_2$, MoSe$_2$ and MoS$_2$, derived from the experiment and modeling. For MoSe$_2$, only one bright trion $X^-$ is measured and drawn.

# Supplemental Material

**Samples and experiment.** Monolayers of the transition metal dichalcongenide (TMDC) semiconductors WSe$_2$, WS$_2$, MoSe$_2$, and MoS$_2$ are mechanically exfoliated from the corresponding bulk single crystalsusing the scotch tape method. Thin flakes (few tens of nm thickness) of hBN are mechanically exfoliated from bulk single crystals (HQ Graphene, Groningen, The Netherlands) with scotch tape as well. A hBN flake, a TMDC monolayer and another hBN flake are successively transferred on a 500 μm thick sapphire substrate using a dry transfer method [1]. Each dry transfer step is followed by annealing at a temperature of 200°C under ambient conditions for 15 minutes. This method ensures good interfacial quality between hBN layers and the TMDC monolayers, which reduces the exciton linewidths, approaching the homogeneous regime.

Optical spectroscopy is performed by keeping the samples in an optical continuous flow He cryostat. For retrieving an absorption spectrum A(λ), we measure reflectance R(λ) and transmittance T(λ) spectra on the same location of the sample under the same experimental conditions. The optical absorption of the sample is calculated as $A(\lambda) = 1 - R(\lambda) - T(\lambda)$. For the reflectivity measurement, light from either a broadband LED source (for WS$_2$ and MoS$_2$) or a tungsten halogen lamp (for WSe$_2$ and MoSe$_2$) is incident on the sample through a 50x long working distance objective lens (numerical aperture NA = 0.55). The reflected light is dispersed using a monochromator with 300 mm focal length, and the spectrum is measured with a peltier-cooled charge coupled device camera. The system response (reference spectrum) is measured by reflecting light from the bare sapphire substrate area close to the hBN/TMDC/hBN structure. Transmission spectra are taken in a similar manner, except that the light is incident on the sample via a achromatric doublet convex lens with 75 mm focal length. It is confirmed using a transfer-matrix analysis that interference effects arising due to reflections from multiple interfaces in the sample are although present in R(λ) and T(λ), they are absent in the absorption $(1 - R(\lambda) - T(\lambda))$ spectra. Therefore, absorption-spectral lineshapes can be directly analysed without taking interference effects into account. For photoluminescence (PL) measurements, a continuous wave laser of 532 nm (2.22 eV) wavelength is focused on a diffraction limited spot. The focused power is noted on the y axis of Figs. 1(e-h) in the main text for the four TMDC materials. Low-temperature spectra ($T = 5K$) for the four samples are presented in Fig. S2.

**Line-shape modeling of the spectra.** Line-shape fitting of the absorption spectra has been performed following Ref. [2]. For the PL spectra, each peak of the PL spectrum is represented by a normalized Lorentzian, and the total spectral intensity $I(E)$ comprising of $n$ peaks is fitted using the following function of energy $E$:

$$I(E) = \frac{1}{2\pi}\sum_{i=1}^{n} \frac{A_i \gamma_i}{(E-E_i)^2 + \left(\frac{\gamma_i}{2}\right)^2} \tag{S1}$$

Here, $A_i, \gamma_i$ and $E_i$ are the intensity, full-width at half-maximum (FWHM) broadening, and the energy of the $i^{th}$ peak, respectively. We start by fitting the lowest temperature data (5 K) first. Here, the exciton ($X^0$), trions ($X_1^-$ and $X_2^-$) and $U^2$ ($U^1$ is not considered in our fitting, since it is enegetically far from the trions, and does not affect their fitting) are clearly resolved. Therefore, their energy positions, linewidth broadenings and the integrated intensities can be obtained with high precision. This is evidenced by small error bars for the intensity of trions in Figs. 2e – 2h of the manuscript. Error bars for $T$ = 5 K spectra for WSe$_2$ and WS$_2$ are smaller than the symbol size ($\pm 1\%$ of the $y$-scale). The parameters obtained in this fit are used as initial parameters for the next (higher) temperature PL data. This leads to an easy convergence of the resulting fit, yielding meaningful values of the fitting parameters. This process is repeated until the spectrum recorded at the highest temperature is reached. For the larger temperatures (T > 75K), we further constrain our fits by assuming a constant separation of 6.5 meV between the two bright trions.

The peaks broaden as temperature rises and start to overlap in the cases of WS$_2$ and WSe$_2$ for $T$ = 75 K onwards. However, following our fitting procedure as described above, the error bars obtained in the fits are reasonably small relative to the trion intensities up to 130 K ($\pm 10\%$ of the y scale at $T$ = 130 K). Taking an example of 1L WSe$_2$, we present the fitting of the $T$ = 5K, 75 K and 150 K PL spectra in Fig. S9. We notice that the intensity of $U^2$ reduces drastically above 75 K, and it does not visibly affect the spectral shape and the fitting of trions after this temperature.

The integrated PL intensities, transition energies and full-width at half-maximum (FWHM) line widths of $U^2, X_1^-, X_2^-$ and $X^0$ obtained from the line-shape fitting are plotted in Fig. S10. The transition energy of the $i^{th}$ quasiparticle peak $E_i$ as a function of temperature in Fig. S10(b) is fitted using the Varshni relation [3,4]

$$E_i(T) = E_{0i} - \alpha_i T^2/(T+\beta) \tag{S2}$$

where $E_{0i}$ is the quasiparticle energy at the absolute zero, and $\alpha_i$ and $\beta$ are the fitting parameters which relate to the dilatation of the lattice and the Debye temperature, respectively [4]. A cumulative fitting was performed for the four curves for the same value of $\beta$. The fitting parameters are given in Table S3. The obtained values of $\alpha = 4.3 \times 10^{-4}$ eV K$^{-1}$ and $\beta = 170$ K for $X^0$ are in excellent agreement with a previous report on 1L WSe$_2$ (i.e. $4.2 \times 10^{-4}$ eV K$^{-1}$ and 170 K, respectively) [3].

The temperature-dependent FWHM linewidth $\gamma_i$ of the $i^{th}$ peak in Fig. S10(c) is fitted using the relationship derived by Rudin et al. [3,5–7]

$$\gamma_i(T) = \gamma_{0i} + \sigma_i T + S_i \frac{1}{e^{\Omega_i/kT} - 1} \tag{S3}$$

Here, $\gamma_{0i}$ is the linewidth at zero temperature, $\sigma_i$ and $S_i$ are the parameters describing coupling of the quasiparticle resonance with acoustic and optical phonons, respectively. $\Omega_i$ is the optical phonon energy. The second term in Eq. (S3) is set to be equal to 0 (i.e. $\sigma_i = 0$), since it is negligibly small compared to the third term. The fitting parameters are provided in Table S3.

The line-shape fitting for the PL spectra of WS$_2$, MoS$_2$, and MoSe$_2$ monolayers is performed the same way. The PL data for $T > 150$ K has not been used for fitting in Fig. 2 (e) – (h), and does not affect (within the fitting error) the derivation of the dark trion energies using our model given by Eq. (1) in the main text.

**Low-temperature absorption and photoluminescence line widths.** The line widths of the trion and exciton resonances lie between 1.8 meV and 6.8 meV (Table S1) at $T = 5$K, which demostrates the excellent quality of the samples. Unresolved resonances are marked with a hyphen. For the transition energies of the resonances, we refer to Table 1 of the main text.

**Model for the transfer of absorption between trions and excitons depending on temperature.** We phenomenologically model the integrated absorption of trions $\alpha_{X^-}$ at different temperatures by a broadened-step Fermi-Dirac-like function (Fig. S5(f)):

$$\alpha_{X^-}(T) = \frac{\alpha_0}{1 + \exp\left(\frac{T - T_0}{T_\gamma}\right)} \tag{S4}$$

where $\alpha_0$ is the integrated absorbance in the limit of $T = 0$ K, and $T_\gamma$ is the width of the distribution around the temperature $T_0$. The fitted parameters are provided in Table S2. The exact values of $T_0$ and $T_\gamma$ depend on i) the complicated interplay of various trion-scattering processes such as phonon-mediated intravalley scattering, intervalley scattering between $K^\pm$ valleys or other points of the Brillouin zone, ii) the curvatures of the bands, and iii) the doping density in the different materials. We use these fitted functions of the integrated absorption for calculating the temperature-dependent behavior of PL intensities of the trions using equation (1) of the main text.

**References.**

[1] A. Castellanos-Gomez, M. Buscema, R. Molenaar, V. Singh, L. Janssen, H. S. J. van der Zant, and G. a Steele, 2D Mater. **1**, 011002 (2014).
[2] A. Arora, T. Deilmann, T. Reichenauer, J. Kern, S. Michaelis de Vasconcellos, M. Rohlfing, and R. Bratschitsch, Phys. Rev. Lett. **123**, 167401 (2019).
[3] A. Arora, M. Koperski, K. Nogajewski, J. Marcus, C. Faugeras, and M. Potemski, Nanoscale **7**, 10421 (2015).
[4] Y. P. Varshni, Physica **34**, 149 (1967).
[5] S. Rudin, T. L. Reinecke, and B. Segall, Phys. Rev. B **42**, 11218 (1990).
[6] Y.-M. He, S. Höfling, and C. Schneider, Opt. Express **24**, 8066 (2016).
[7] L. Zhang, R. Gogna, W. Burg, E. Tutuc, and H. Deng, Nat. Commun. **9**, 713 (2018).
[8] T. Deilmann and K. S. Thygesen, Phys. Rev. B **96**, 201113 (2017).
[9] M. Drüppel, T. Deilmann, P. Krüger, and M. Rohlfing, Nat. Commun. **8**, 2117 (2017).

**Table S1.** Low-temperature ($T = 5$K) FWHM line widths of excitons (X) and trions ($X_1^-$, $X_2^-$) derived from PL and absorption spectra of the four TMDC monolayer crystals encapsulated in hBN

| Material (encapsulated with hBN) | Full-width at half-maximum line width (meV) | | | | | |
|---|---|---|---|---|---|---|
| | $X_1^-$ | | $X_2^-$ | | $X^0$ | |
| | Abs. | PL | Abs. | PL | Abs. | PL |
| 1L WSe$_2$ | 4.0 | 3.2 | 5.0 | 4.8 | 4.4 | 6.8 |
| 1L WS$_2$ | 2.0 | 2.6 | 2.6 | 2.1 | 5.5 | 6.6 |
| 1L MoSe$_2$ | 4.0 | 5.4 | – | – | 4.0 | 6.7 |
| 1L MoS$_2$ | 2.3 | 1.8 | 4.4 | – | 4.0 | 2.4 |

**Table S2.** $T_0$ and $T_\gamma$ parameters obtained after fitting Eq. (S1) to data in Figs. 2(a) to 2(d).

| Material (encapsulated with hBN) | $T_0$ (K) | | $T_\gamma$ (K) | |
|---|---|---|---|---|
| | $X_1^-$ | $X_2^-$ | $X_1^-$ | $X_2^-$ |
| 1L WSe$_2$ | 70 | 75 | 16 | 17 |
| 1L WS$_2$ | 5 | 5 | 25 | 30 |
| 1L MoSe$_2$ | 100 | – | 46 | – |
| 1L MoS$_2$ | 140 | 50 | 17 | 23 |

**Table S3.** Fitting parameters of the temperature-dependent pholuminescence data of hBN-encapsulated 1L WSe$_2$ in Fig. S10(b and c)

| Parameter | $U^2$ | $X_1^-$ | $X_2^-$ | $X^0$ |
|---|---|---|---|---|
| Fitting parameters for Fig. S10(b) (using Eq. (S2)) | | | | |
| $E_0$ (eV) | 1.678 | 1.692 | 1.698 | 1.728 |
| $\alpha$ ($\times 10^{-4}$ eV K$^{-1}$) | 3.8 | 4.6 | 4.5 | 4.3 |
| $\beta$ (K) | 170 | | | |
| Fitting parameters for Fig. S10(c) (using Eq. (S3)) | | | | |
| $\gamma_0$ (meV) | 5.3 | 4.0 | 5.5 | 6.3 |
| $S$ (meV) | 11.7 | 32.7 | 66.5 | 34.6 |
| $\Omega$ (meV) | 5.5 | 12.0 | 23.4 | 19.7 |

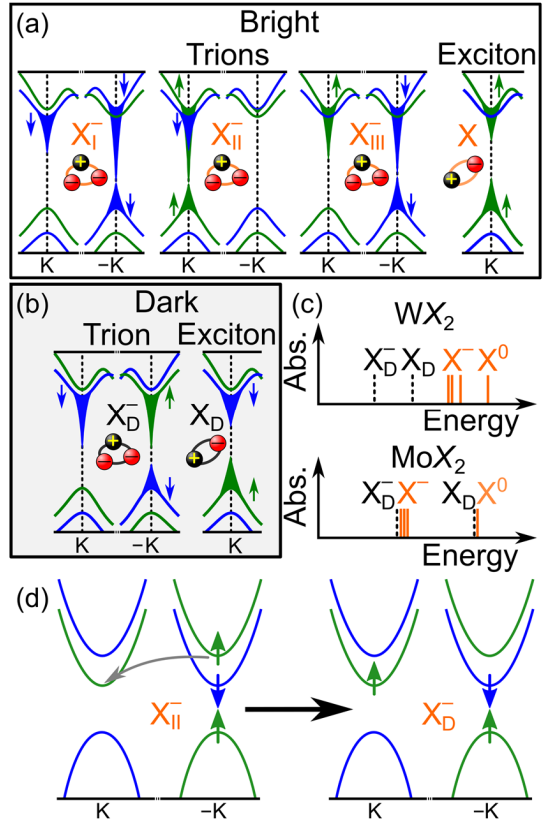

**Figure S1.** (a) k-space representation of the three possible types of bright trions, $X_I^-$, $X_{II}^-$ and $X_{III}^-$ (not to be confused with labels of experimentally observed trions $X_1^-$ and $X_2^-$ in the main text) and the bright exciton $X^0$ in the monolayers of W-based semiconducting transition metal dichalcogenides where bands are colored blue (down spin) and green (up spin). The bands participating in the creation of an excitation are represented by elongated filled cones. The third trion ($X_{III}^-$) has both electrons in the upper conduction bands, and is expected to have low oscillator strength, explaining its absence in the experiments. (b) is similar to (a) for the dark excitations (see Ref. [8] for details). Excitations similar to those shown in (a) and (b) with opposite spin and valley character are also possible but not displayed for brevity. (c) Schematic drawing showing the relative energetic positions of the dark excitations (black dashed) with respect to the bright (orange) absorption lines in W- and Mo-based TMDC monolayers. For W-based materials, dark states $X_D$ and $X_D^-$ are lower in energy than the bright states $X^0$ and $X_D$, in contrast to Mo-based materials, where the dark exciton (trion) is in close proximity to the bright exciton (trion). In the presence of hBN layers encapsulating the TMDC monolayer, a large band gap renormalization and a small red-shift of the excitations occurs [9]. However, the effect of surrounding hBN on the relative separation between bright and dark trions (when compared to vacuum around the monolayer) is less than 10 meV, which is within our numerical accuracy (d) Bright trions can convert to dark trions and vice versa via intervalley scattering processes of the excess carriers, without the requirement of a spin flip. An example of one such process is shown.

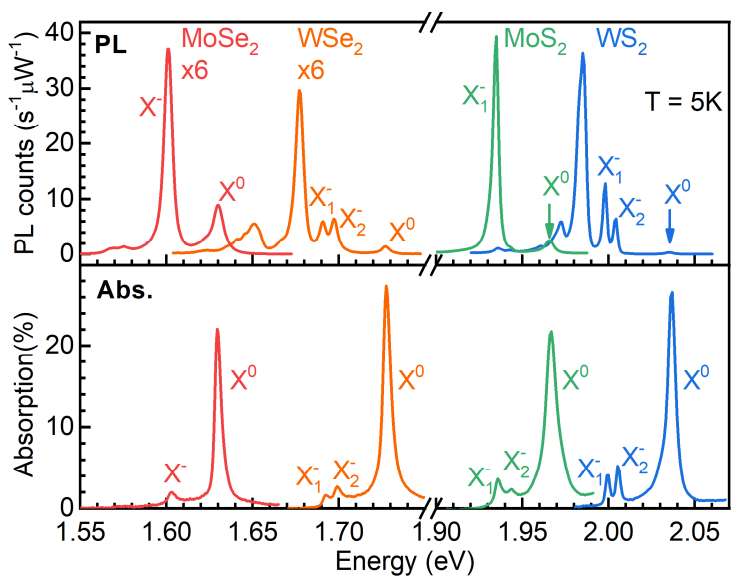

**Figure S2. Low-temperature excitonic spectra.** (Top) Low-temperature ($T = 5K$) photoluminesence spectra of MoSe$_2$, WSe$_2$, MoS$_2$, and WS$_2$ monolayers encapsulated in hBN. A 532 nm CW laser with diffraction-limited focus on the sample and a focused power of 500 µW, 200 µW, 60 µW, and 80 µW, respectively is used for the PL measurements. PL counts are divided by the accummulation time as well as power (in µW) to directly compare the four cases. MoSe$_2$ and WSe$_2$ spectra are amplified by factors of 6 for clarity. (bottom) Corresponding measured absorption spectra $A(\lambda) = 1 - R(\lambda) - T(\lambda)$ for the four samples.
55

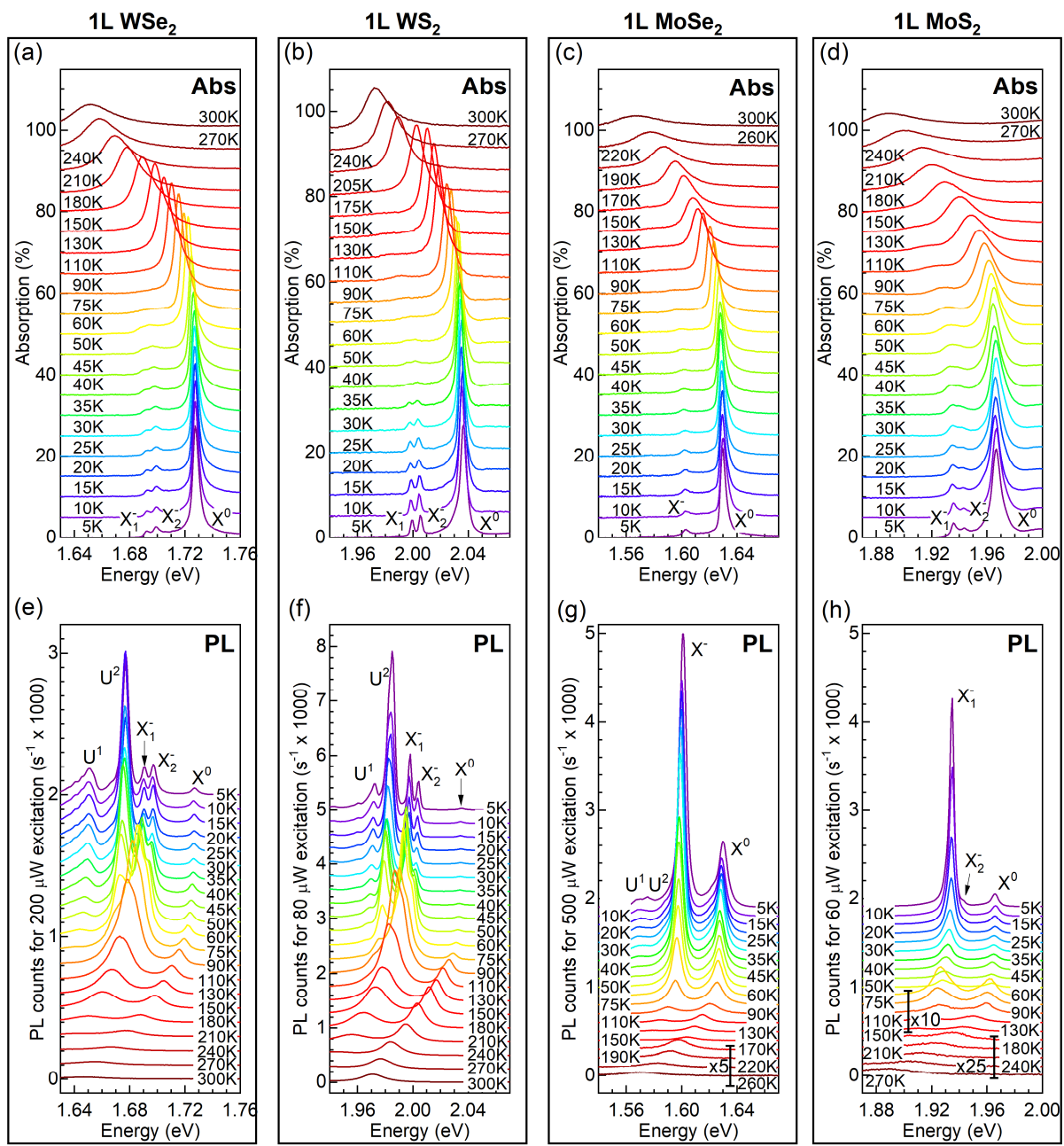

**Figure S3. Absorption and PL spectra for all temperatures.** Optical absorption spectra of hBN-encapsulated (a) WSe$_2$, (b) WS$_2$, (c) MoSe$_2$, and (d) MoS$_2$ monolayers on sapphire substrate, as a function of temperature $T = 5 - 300$ K. (e) to (h) photoluminescence spectra of the four materials as a function of temperature. The spectra are vertically shifted for clarity. Low PL intensities in (g) and (h) are amplified by factors mentioned with the spectra.

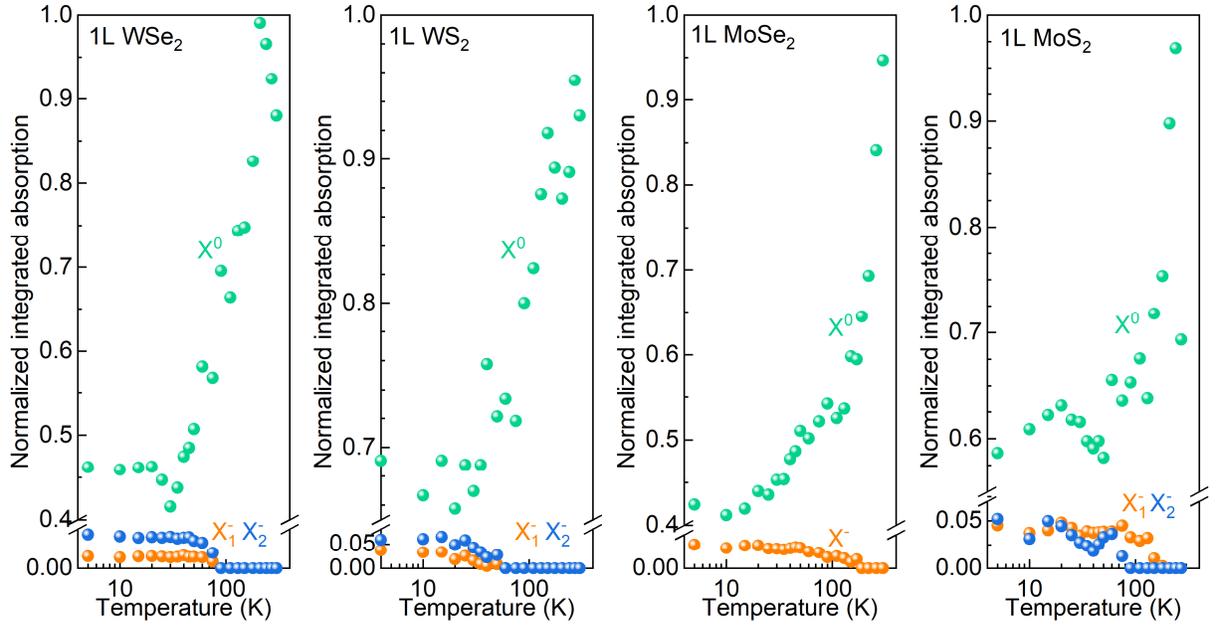

**Figure S4. Transfer of integrated absorption between excitons and trions.** Temperature-dependent normalized integrated absorption (area under the absorption peak) derived from the line-shape analysis of the absorption spectra of monolayers of WSe$_2$, WS$_2$, MoSe$_2$, and MoSe$_2$ encapsulated in hBN are plotted. For all four TMDC monolayers, the integrated absorption of the neutral exciton rises, while that of charged excitons reduces. A Fermi-Dirac distribution based model explains the observed trend as discussed in the main text.

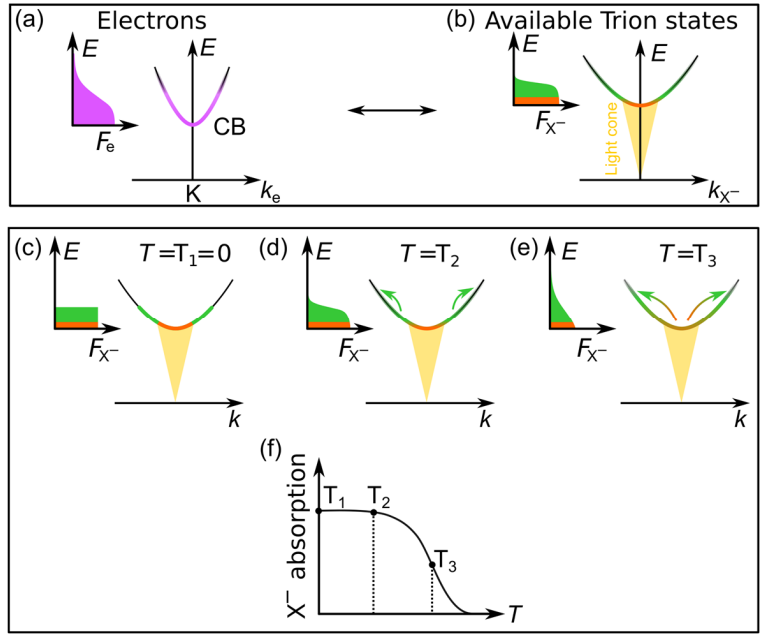

**Figure S5. Reduction of integrated absorption of trions with temperature.** a) Thermal distribution of electrons (purple) in the conduction band (CB), which is linked to the available trion states (b), in momentum space. The F-D distribution functions in the two cases are related by $F_{X^-}(T) = F_e(T\, m^*_{X^-}/m^*_e)$, i.e. the temperature is rescaled by the mass ratio of the quasiparticles. The optically active (within the light cone) and inactive trion states are marked in orange and green, respectively. Because of a larger mass, the trion band has a smaller curvature. (c) At $T = T_1 = 0$, the trions have plenty of optically active states available, resulting in a large integrated absorption, as shown in (f). (d) For $T = T_2 > T_1$, carriers outside the light cone redistribute to higher energies. However, the number of available trion states within the light cone is largely unaffected, and the integrated absorption is similar to that in (c). (e) For even higher temperature ($T = T_3$), the probability of creating trions within the light cone reduces due to the redistribution of carriers to higher energies, and the integrated absorption is substantially reduced, as depicted in (f).

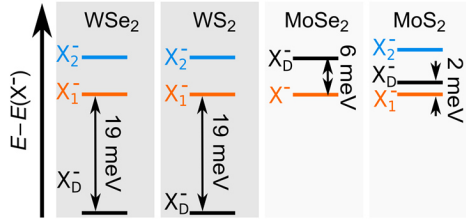

**Figure S6.** Relative energies of the dark trion $X_D^-$ with respect to the bright trion $X_1^-$, for hBN-encapsulated monolayers of WSe$_2$, WS$_2$, MoSe$_2$ and MoS$_2$, derived from the experiment and modeling. The second bright trion $X_2^-$ is also marked. For MoSe$_2$, only one bright trion $X^-$ is measured and drawn.

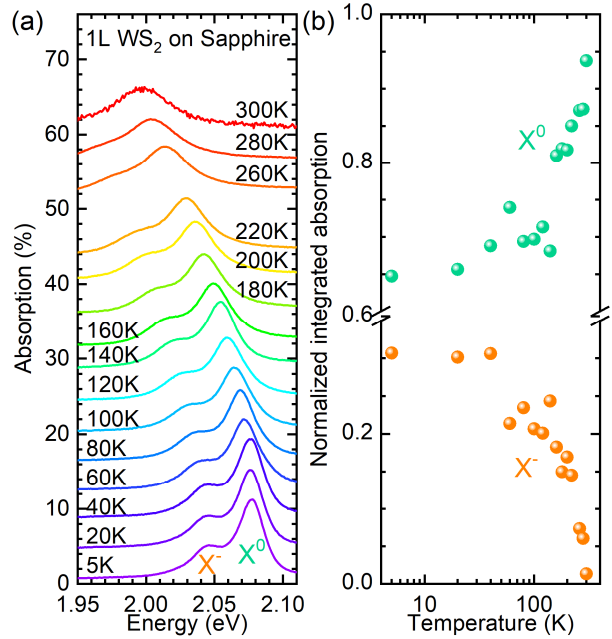

**Figure S7. Temperature-dependent absorption of 1L WS$_2$ on sapphire substrate.** (a) Optical absorption spectra of a WS$_2$ monolayer (not encapsulated) on sapphire substrate, as a function of temperature $T = 5 - 300$ K (temperatures are indicated in the plot). The spectra are shifted along the y axis for clarity. Absorption lines correspond to the trion $X^-$ and the neutral exciton $X^0$. $X^-$ consists of two overlapping peaks $X_1^-$ and $X_2^-$, which are not resolved, because the monolayer is not encapsulated between hBN layers. b) Integrated absorption (area under the absorption peak) of $X^-$ and $X^0$ as a function of temperature. Similar to the hBN-encapsulated 1L WS$_2$ sample in Fig. 2b of the main text and Fig. S4, the integrated absorption of trions decreases, while it increases for the neutral excitons with rising temperature.

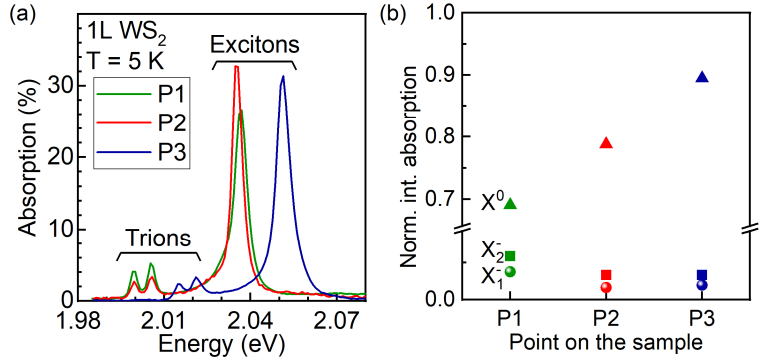

**Figure S8. Absorption spectra measured at different locations of the hBN-encapsulated 1L WS$_2$ sample.** (a) shows absorption spectra of an hBN-encapsulated WS$_2$ monolayer, measured on different positions (P1 to P3) of the sample. The energies of the resonances at P3 are strongly different different compared to P1 and P2, possibly due to different dielectric conditions, doping, and strain. In (b), the integrated absorption of the two trions and the neutral exciton are plotted for the three locations in (a). One finds that a lower integrated absorption of the trions results in a higher integrated absorption of the neutral exciton. This is in qualitative agreement with our model in Fig. S5, where a larger occupation of momentum space by the trions inhibits the creation of neutral excitons. The data presented in the main text of this manuscript is measured at P1.

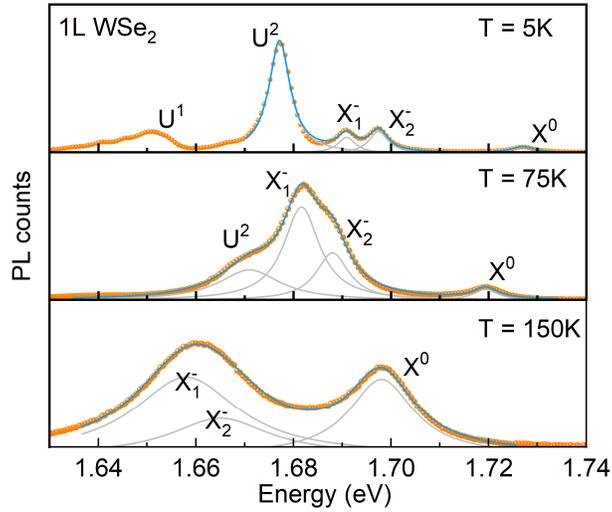

**Figure S9. Line-shape modeling of hBN-encapsulated 1L WSe$_2$ PL.** (a) Examples of line-shape modeling (blue solid lines) of the PL spectra (orange spheres) of hBN-encapsulated WSe$_2$ for $T = 5\,\text{K}, 75\,\text{K}$, and $150\,\text{K}$ using Eq. (S1). For 5 K and 75 K spectra, four Lorentzians are considered, while for 150 K, only three Lorentzians are sufficient. The PL peaks are marked as $U^1, U^2, X_1^-, X_2^-$, and $X^0$ as explained in the main text.

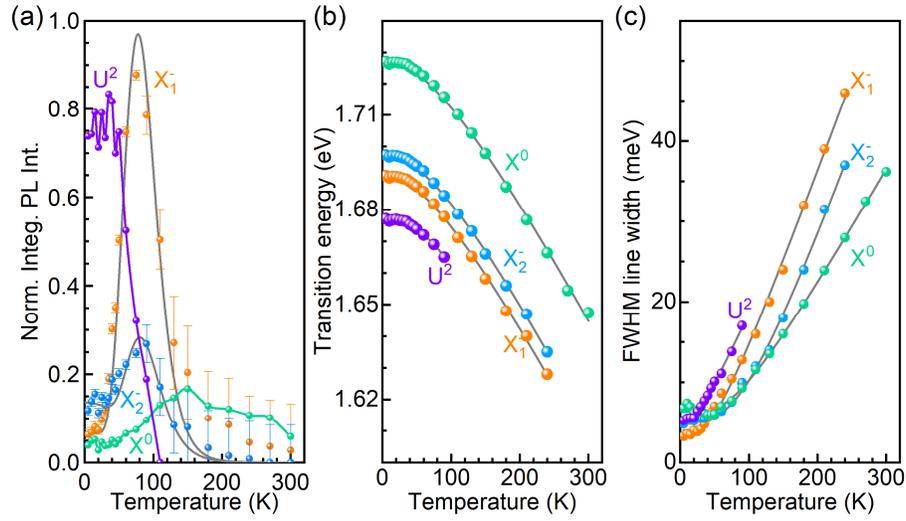

**Figure S10. Fitting parameters (Eq. (S1)) of hBN-encapsulated 1L WSe$_2$ PL.** (a) Integrated PL intensities (normalized), (b) transition energies, and (c) full-width at half-maxima (FWHM) of $U^2$, $X_1^-$, $X_2^-$, and $X^0$ (spheres) derived from the line-shape modeling of the PL spectra as a function of temperature using Eq. (S1). Solid grey lines in (a) to (c) represent the modeled curves to the data as explained as follows. Modeled curves in (a) are reproduced from Fig. 2(e) of the main text. No modeling is performed for the PL intensities of peaks $U^2$ and $X^0$ where, the solid lines joining the experimental points are a guide to the eye. Transition energies in (b) are modeled using Eq. (S2), while the FWHM line widths are modeled using Eq. (S3). The parameters obtained are provided in Table S3.